\documentclass[prl,preprint,showpacs]{revtex4}
\usepackage{epsf}
\usepackage{psfig}
\usepackage{graphicx}
\begin{document}

\title{Optimal operating conditions and characteristics of acetone/CaF$_2$ detector for inverse photoemission spectroscopy}

\author{S. Banik, A. K. Shukla, and S. R. Barman$^*$}

\affiliation{UGC-DAE Consortium for Scientific Research,
Khandwa Road, Indore, 452017, Madhya Pradesh, India.}

\begin{abstract}
Performance and characteristics of a band-pass  photon detector using acetone gas and CaF$_2$ window (acetone/CaF$_2$) have  been studied and  compared with an ethanol/MgF$_2$ detector. The optimal operating conditions are found to be 4~mbar acetone pressure and 745$\pm$20~V anode voltage. The count rate obtained by us is about a factor of 3 higher than what has been reported earlier for the acetone detector. Unlike other gas filled detectors, this detector works in the proportional region with very small dead time (4~$\mu$sec). A detector band-pass of 0.48$\pm$0.01~eV FWHM is obtained. 
\end{abstract}
\maketitle

\baselineskip=36pt
Inverse photoemission spectroscopy (IPES) has emerged as a powerful technique to study the unoccupied electron states.\cite{Dose83,Smith88,Johnson90} Gas filled band-pass Geiger-M\"{u}ller (GM) type counters are generally used for photon detection in IPES because of their high efficiency, low cost and simple design.
GM detectors have played an important role in IPES, since their first use by Dose and co-workers.\cite{Denninger79} Acetone/CaF$_2$ detector was first used by Funnemann and Merz who reported a band-pass of 0.4~eV FWHM.\cite{Funnemann85} But detailed characteristics of the acetone/CaF$_2$ detector has not been reported in literature. 

The design of our detector and the electron source is similar to those already reported in literature.\cite{Denninger79,Funnemann85,Stoffel85} The fill-gas pressures are measured by convectron gauge from Granville Phillips, USA (model 375) as uncorrected apparent N$_2$ pressures. In order to determine the optimal operating conditions,  we show in Fig.1a the count rate ($C$) of acetone/CaF$_2$ at the Fermi edge of polycrystalline Ag as functions of acetone gas pressure ($p$) and anode voltage ($V_A$). 
$p$ has been varied from 0.3 to 9~mbar, while $V_A$ has been varied from 300~V till just below the breakdown voltage ($V_B$). Maximum $C$ is observe in a region starting from about 3~mbar and 650~V to 6~mbar and 1050~V (violet/indigo region in Fig.1a). As $p$ is increased, anode voltage  corresponding to maximum $C$ ($V_M$) increases. Above 6.5 mbar, $V_M$ increases sharply and the violet region shifts to 1400~V. In Fig.1b, we show $C$ as a function of $V_A$ for different $p$. The choice of  optimum operating conditions, besides $C$, also depends on the stability of the detector. We observed that $C$ is similar between 3 to 6.5 mbar (Fig.1). But at $p$= 5-6.5~mbar, $V_M$  is quite large ($>$1000~V) and this can cause operational problems with no gain in $C$. In contrast, at lower $p$ although $V_M$ is relatively less ($\approx$650~V), a small increase in the voltage beyond $V_M$ could result in breakdown. For example, below $p$= 2~mbar (see 0.9 and 1.9 mbar plots in Fig.1b) $C$ is less and $V_B$ and  $V_M$ almost coincide.  Hence, at this operating condition, the acetone detector could be quite unstable and breakdown can occur. At 2.8~mbar $V_M$ is similar to 1.9~mbar, while $V_B$ increases to about 700~V from 650~V. At 4 mbar, the maximum $C$ is obtained over a voltage range given by 745$\pm$20~V. 
The difference between $V_B$ and $V_M$ increases with $p$. $V_B$ shows a steep increase for $p$$>$ 2~mbar, as shown in Fig.1b (inset).  This indicates that at higher $p$ the detector is more stable. {\it We choose the optimal operating condition where $C$ is maximum and $V_B$ is considerably higher than $V_M$ at $p$=~4~mbar, $V_M$=~745$\pm$20~V with $V_B$ about 50~V higher.} We find that the operating condition given in Ref.\cite{Funnemann85} (0.75 mbar and 650~V) is in the unstable region where $V_M$ and $V_B$ are very close. This is possibly the reason why the acetone detector is generally considered to be unstable. 

We have studied the stability of $C$ as a function of time during continuous operation. The detector requires about 30-40~minutes to stabilize during which $C$ decreases by 15-25\%. We report the stablized $C$ at the optimal operating conditions (4 mbar, 745$\pm$20~V) to be  300$\pm$25 counts/$\mu$A~sec~sr. This is lower than $C$  shown in Fig.1 where it is  not stabilized.  
Maximum $C$ that we obtain is about 3 times higher than what was reported earlier (105 counts/$\mu$A~sec~sr).\cite{Funnemann85} A direct comparison can be made with Ref.\cite{Funnemann85}, since in both cases $C$ is measured at the Fermi edge of polycrystalline Ag, which is the main reason why we have used Ag as our test sample. Although $C$ is considerably less, it is possible to operate the detector at 0.75 mbar by carefully increasing $V_A$.
Generally, Ar or He is used as multiplier gas in the detector to facilitate the Geiger discharge.\cite{Knoll} However, we find Ar to be detrimental for the  acetone detector since $C$  decreases rapidly with introduction of Ar.  

From the shape of the pulse height distribution, it was inferred that the acetone detector works in the proportional region,\cite{Funnemann85} while ethanol detector works in the GM region.\cite{Duffin02} We find that the pulse height distributions of acetone and  ethanol detectors exhibit similar exponential behavior, the acetone distribution being slightly broader. So it is difficult to definitively infer from the pulse height distribution about the region of operation (proportional or GM). A direct way to ascertain this is by determining the average pulse height ($\mu$)  as a function of $V_A$.\cite{Knoll} 
$\mu$ for the acetone detector at the output of amplifier (Fig.2a) shows an increase around 450~V (onset voltage), after which it exhibits almost a plateau. 
An increasing trend is observed above 630~V, while above   750~V (which corresponds to $V_M$) $\mu$ increases drastically. The standard deviation  of the pulse height ($\sigma$) gradually increases with $\mu$ and exhibits a large increase beyond $V_M$ (Fig.2a). Above $V_M$, the occurrence of  large pulses ($>$12~V) becomes more frequent, which probably leads to the  increase in $\sigma$ and is a fore-warning for detector breakdown. Comparing the variation of $C$   and $\mu$ (Fig.1b and 2a), we find that above $V_M$, $\mu$ increases but $C$  decreases till the breakdown is reached ($V_B$=~800~V).  The increase in $\mu$ above 630~V indicates the onset of the proportional region of operation.\cite{Knoll} {\it Thus, under optimized operating conditions (4 mbar, 745$\pm$20 volts), the acetone/CaF$_2$ detector works in the proportional region.} Above 750~V, $\mu$ increases substantially indicating that the detector is going into the Geiger region of operation, but here $C$  actually decreases. Moreover, $\sigma$ becomes larger than $\mu$ and the detector becomes unstable.  A possible reason that prevents the operation of acetone detector in the Geiger region could be fragmentation. It has been reported that  photon induced fragmentation of acetone starts at h$\nu$=~10.52~eV, where acetone (CH$_3$COCH$_3$) fragments into CH$_3$CO$^+$, CH$_3$ and an electron.\cite{Trott78} 
In the  region above $V_M$, $C$ decreases with increasing voltage (Fig.1b), indicating the start of  fragmentation process. At $V_B$,  almost complete fragmentation occurs, the fragments have to be pumped out and the detector needs  refilling. However, for   normal operation at optimal conditions, breakdown does not occur. 

From Fig. 2a, $\mu$ is about 1~V at $V_M$. Using this and considering the electronic gain, the number of electrons in each charge pulse is estimated to be about 10$^4$. In contrast, we find that the ethanol detector gives 10$^8$ electrons/pulse. Thus, the charge pulses in the acetone detector are four orders of magnitude less than in the ethanol detector. The number of electrons in an avalanche is related to the region of operation of the detector. For proportional counters, this number is 10$^2$ to 10$^4$, while for GM counters it is 10$^8$ to 10$^9$.\cite{Knoll} This shows that while the ethanol detector works in the GM region, the acetone detector operates in the proportional region. 

We show in Fig.2b(v) the pulses at the amplifier output of the ethanol detector, where the first pulse has been triggered at time t=~0.  This gives the behavior of the pulses up to 500~$\mu$sec after the appearance of the t=~0 pulse by repeated superposition for 60 minutes in the envelope mode of the CRO. The electron source is kept at a fixed kinetic energy. We obtain an envelope comprising of the overlapping pulses after different t=~0 pulses. Clearly, no pulses are observed for about 100~$\mu$sec, while the region above 100~$\mu$sec is completely filled up with pulses so that the individual pulses cannot be distinguished. Absence of pulses between 0 and  100~$\mu$sec shows that after the $t$=~0 pulse the detector cannot count for 100~$\mu$sec, which is the dead time.\cite{Knoll} After 100~$\mu$sec, the pulse height increases exponentially, and the time constant is known as the recovery time. By fitting the envelope with an exponential function as in Ref.\cite{Duffin02}, we find the recovery time to be 126~$\mu$sec [Fig.~2b(v)]. This value is somewhat smaller than in Ref.\cite{Duffin02}, which could possibly arise from differences in the detector geometry, pulse height {\it etc.}  

Interestingly, in case of acetone/CaF$_2$, we find a completely different behavior [Fig.2b(i-iv)]. Immediately (\verb+"+0\verb+"+ minutes) after  starting the pulse recording, shows almost no pulses after the t=~0 pulse [Fig.2b(i)]. After 5 minutes, there are pulses of different heights over the whole range. After 60 minutes, more number of  pulses with larger height are observed as they superpose on the smaller pulses. Since the amplifier output has a cut-off of maximum 12~V, pulses higher than that are truncated. After 150 minutes [Fig.2b(iv)], the occasional large pulses ($\ge$12~V) smear the whole region. If Figs.2b(i-iv) are compared with Fig.2b(v), it is immediately clear that there is no dead time or recovery time in the acetone detector. In this case, the typical width of a pulse, which is about 4~$\mu$sec, could be taken as an estimate of the dead time. Such small dead time shows that the acetone detector works in the proportional region.\cite{Knoll}

We show the IPES spectrum of polycrystalline silver  recorded using  
acetone detector without any dead time correction. Dead time correction is not required since it is very small for acetone detector.  The  spectrum is in good  agreement with literature (Fig.3a).\cite{Funnemann85,Reihl84} 
We find that the spectral shape remains same for spectra recorded at different $p$ and $V_A$.  So, the Ag spectrum in Ref.\cite{Funnemann85} recorded at $p$=~0.75 mbar and 650~V agrees with our spectra recorded at $p$=~4~mbar and 745~V. 
In order to estimate the resolution of the Ag spectrum, we have collected data in the near E$_F$ region with small step size, 0.02~eV (Fig.3b). To obtain a parameter dependent fitting function, a spectral function $P(E)$ is  multiplied by the Fermi function $F(E)$ and this is convoluted with an instrumental resolution function $T(E)$. $T(E)$ represents  the broadenings due to the electron source and the detector band-pass function.  $T(E)$ is generally approximated by a Gaussian.\cite{Dose83,Funnemann85} Since the Ag DOS near $E_F$ is reasonably flat, $P(E)$ is approximated by a constant function. Least square error minimization is performed using Levenberg-Marquardt algorithm.  The Gaussian width, position and intensity of $F(E)$ are freely varied. The residual in the bottom of Fig.3b shows the good quality of the fit. 
We determine the overall instrumental resolution to be 0.55$\pm$0.01~eV, which is the FWHM of the Gaussian function $T(E)$ (Fig.3b), in excellent agreement with Ref.\cite{Funnemann85}. 
This shows that there is no significant change in resolution for different operating conditions. Moreover, it also shows that the procedure followed in  Ref.\cite{Funnemann85}  for obtaining the resolution\cite{Dose83} gives similar result as our least square fitting approach. 
We determine the position of the Fermi level (shown by vertical line) to be 8.18$\pm$0.001~eV. To determine the detector band-pass, we estimate the FWHM of the electron source to be about 0.26~eV (1200~K) assuming a Maxwellian distribution. Subtracting this  from the overall resolution, we obtain the detector band pass to be 0.48$\pm$0.01~eV FWHM. Thus, the band-pass of the acetone detector is larger than the ethanol detector (0.35~eV). From our result, it seems that the acetone/CaF$_2$ band-pass of 0.4~eV reported in Ref.\cite{Funnemann85} using data from literature\cite{Trott78} is an under-estimation. This is evident from Fig.5 of Ref.\cite{Funnemann85}, where the experimentally determined resolution function (curve A) has larger FWHM (0.55 eV) compared to the convolution of estimated detector function with the electron source width (curve B).
 
This work is funded by D.S.T. under project no. SP/S2/M-06/99. K. Horn, D. D. Sarma, A. B. McLean, and C. Biswas are thanked for useful discussions. B. A. Dasannacharya and  A. Gupta are thanked for encouragement. V. Bhatnagar, S. C. Das and P. Ramshankar are thanked for technical help.

\newpage
\noindent$^*$Electronic mail: barman@udc.ernet.in

\newpage
\thispagestyle {empty}
\noindent {\large Figure Captions :}\\

\noindent Figure 1.~(a) (Color online) Count rate ($C$) of the acetone detector as functions of acetone pressure ($p$) and the anode voltage ($V_A$). In rainbow color scheme, red contour represents zero $C$, while violet contour represents maximum $C$ (380$\pm$25 counts/$\mu$A~sec~sr). (b) $C$ versus $V_A$; $V_M$ and $V_B$ are indicated by arrow and tick, respectively; inset shows $V_B$ as a function of $p$.
 
\noindent Figure 2. (a) Average ($\mu$) and standard deviation ($\sigma$) of the pulse height of acetone detector as a function of $V_A$ at $p$=~4~mbar. (b) Pulses  at amplifier output after triggering the first pulse at t=~0  for  (i-iv) acetone/CaF$_2$  and (v) ethanol/MgF$_2$ detectors. In (v), the exponential fit is shown slightly shifted along the vertical axis. The time duration (in minutes) of data collection is shown in each case.

\noindent Figure 3. (a)Inverse photoemission spectrum of Ag. (b) Near $E_F$ region of the Ag spectrum (open circles),  fitted curve (thick solid  line), Fermi function ($F(E)$, dashed line),  instrumental Gaussian function ($T(E)$, thin solid line) and the residual of fitting are shown.


\begin{thebibliography}{}
\bibitem{Dose83} V. Dose, Prog. Surf. Sci. {\bf13}, 25 (1983).
\bibitem{Smith88} N. V. Smith, Rep. Prog. Phys. {\bf51}, 1227 (1988).
\bibitem{Johnson90} P. D. Johnson and S. L. Hubert, Rev. Sci. Instrum. {\bf61}, 2277 (1990).
\bibitem{Denninger79} G. Denninger, V. Dose and H. Scheidt, Appl. Phys. {\bf18}, 375 (1979).
\bibitem{Funnemann85} D. Funnemann and H. Merz, J. Phys. E: Sci. Instrum. {\bf19}, 554 (1985).
\bibitem{Stoffel85} N. G. Stoffel and P. D  Johnson, Nucl.  Instrum. and  Meth. in Phys. Res. A {\bf 234}, 230 (1985).
\bibitem{Knoll} G. F. Knoll, {\it Radiation Detection and Measurement} (Wiley, New York, 1989).
\bibitem{Duffin02} J. A. Lipton-Duffin, A. G. Mark, and A. B. McLean, Rev. Sci. Instrum. {\bf73}, 3149 (2002).
\bibitem{Trott78} W. M. Trott, N. C. Blais, and E. A. Walters, J. Phys. Chem. {\bf69}, 3150 (1978); J. L. Campbell and K. W. D. Ledingham, Brit. J. Appl. Phys., {\bf 17}, 769 (1966).
\bibitem{Reihl84} B. Reihl and R. R. Schlittler, Phys. Rev. B {\bf 29}, 2267 (1984).
\end{thebibliography}
\end{document}